\documentclass[amsmath,amsfonts,twocolumn,superscriptaddress,prx,longbibliography]{revtex4-1}
\usepackage{graphicx}
\usepackage{epsfig}
\usepackage{bm}
\usepackage{dcolumn}
\usepackage{amsmath}
\usepackage{amssymb}
\usepackage{color}
\usepackage[varg]{txfonts}

\DeclareMathOperator\arccosh{arccosh}

\begin{document}

\title{Thermodynamic properties of nodal superconductors close to a magnetic quantum critical point}

\author{Jaglul Hasan}
\affiliation{Department of Physics, University of Wisconsin--Madison, Madison, Wisconsin 53706, USA}

\author{Maxim Dzero}
\affiliation{Department of Physics, Kent State University, Kent, Ohio 44242, USA}

\author{Maxim Khodas}
\affiliation{Racah Institute of Physics, Hebrew University of Jerusalem, Jerusalem 91904, Israel}

\author{Alex Levchenko}
\affiliation{Department of Physics, University of Wisconsin--Madison, Madison, Wisconsin 53706, USA}

\begin{abstract}
In this work we study thermodynamic manifestations of the quantum criticality in multiband unconventional superconductors. As a guiding example we consider the scenario of magnetic quantum critical point in the model that captures superconductivity coexistence with the spin-density wave. We show that in situations when the superconducting order parameter has incidental nodes at isolated points, quantum magnetic fluctuations lead to the renormalization of the relative $T$-linear slope of the London penetration depth. This leads to the nonmonotonic dependence of the penetration depth as a function of doping and the concomitant peak structure across the quantum critical point. In addition, we determine contribution of magnetic fluctuations to the specific heat at the onset of the coexistence phase. Our theoretical analysis is corroborated by making a comparison of our results with the recent experimental data from the low-temperature thermodynamic measurements at optimal composition in BaFe$_2$(As$_{1-x}$P$_x$)$_2$.    
\end{abstract}

\date{February 9, 2022}

\maketitle

\section{Introduction}

In superconductors with unconventional symmetry of the pairing order parameter, nodal structure of the gap leads to distinct 
power-law temperature dependencies, $\propto T^a$, of various thermodynamic properties and electronic kinetic coefficients \cite{Mineev-Samokhin,Sigrist-Ueda}. This signature behavior is in sharp contrast to that found in conventional $s$-wave superconductors where these quantities exhibit thermally activated exponential behavior, $\propto e^{-\Delta/T}$, determined by the spectral energy gap $\Delta$. Moreover, the power exponent distinguishes between the types of the nodes. For instance, in the case of the heat capacity $a=3$ if the symmetry enforced gap structure has point (or first-order) nodes at isolated points, while $a=2$ if the gap has line-nodes or second-order nodes at isolated points. Similar conclusions can be drawn for the other thermodynamic quantities including London penetration depth and Knight shift, as well as the kinetic coefficients such as electronic thermal conductivity and ultrasound attenuation (see, for example,  Refs. \cite{Mineev-Samokhin,Sigrist-Ueda,ProzorovKogan,Arfi,Moreno,Joynt} and references therein).    

In strongly correlated materials, including families of cuprates, iron-pnictides, and heavy fermion systems, superconducting instability often competes with some other form of long-range electronic ordering, such as spin or charge density wave orders; see Refs. \cite{Scalapino,Chubukov,Proust,Qi-RMP20}. 
At least in some parameter range of the phase diagram, the interplay between the corresponding interactions may lead to energetically stable coexistence of superconductivity and other order. The line of second order phase transition separating the pure state of unconventional superconductivity and mixed phase of superconductivity coexisting with, e.g., spin-density wave order terminates at the quantum critical point (QCP). Thus far accumulated experimental evidence suggests that quantum criticality manifests itself with various anomalies in thermodynamic response functions (see \cite{QCP-Review} for the detailed review and references therein). Notable recent examples include observations of (\emph{i}) nonmonotonic discontinuity of the specific heat jump that peaks at the QCP \cite{Hardy,Carrington,Grinenko}, (\emph{ii}) a sharp peak of the low-temperature magnetic penetration depth $\lambda$ occurring at the fine-tuned material composition \cite{Hashimoto1,Hashimoto2,Auslaender,Joshi}, and (\emph{iii}) related anomalies in the lower $H_{c1}$ and upper $H_{c2}$ critical fields \cite{Putzke}. These anomalous properties persist into the normal state, most profoundly in the form of linear-in-$T$ Planckian resistivity observed in various materials at the optimal doping \cite{Mackenzie,Analytis,Shekhter,Taillefer}.   

\begin{figure}[t!]
  \centering
 \includegraphics[width=3.25in]{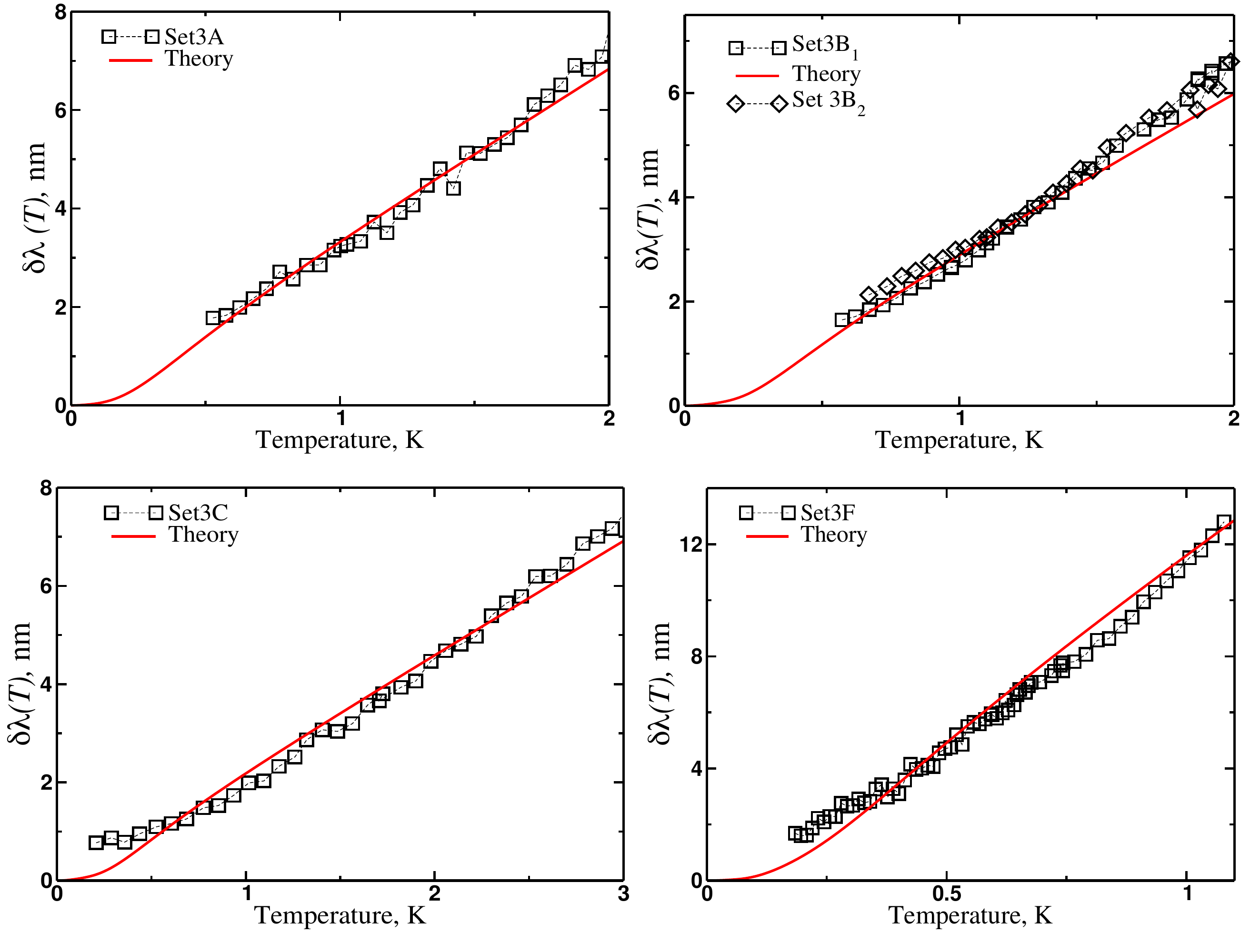}
 \caption{Fits for the relative change of the penetration depth, $\delta\lambda(T)$(nm), measured in BaFe$_2$(As$_{1-x}$P$_x$)$_2$ at low temperatures and plotted for different compositions in the range $x=0.27-0.64$. For this compound the QCP is located at $x_c=0.3$. The labeling of the panels corresponds to that of Fig. 3 from Ref. \cite{Hashimoto1}. For panel B data for two different samples are shown on top of each other distinguished by squares $\Box$ and rhombus $\Diamond$.}\label{Fig1}
\end{figure}

In this work we consider a model of a multiband metal that captures coexistence of superconductivity (SC) with the spin-density wave (SDW) state and realizes a particular example of a magnetic quantum critical point in a superconducting state. We show that existence of incidental nodes in the superconducting order parameter on some Fermi surfaces leads to linear in temperature dependence of the relative change in the penetration depth, $\delta\lambda=\lambda(T)-\lambda(0)\propto T$. In addition, quantum fluctuations associated with the emergent SDW order at the onset of QCP lead to noticeable renormalizations that change the slope of $T$-linear behavior that peaks at QCP. In particular, this explains the main observations reported in Ref. \cite{Hashimoto1} concerning the doping and temperature evolution of the penetration depth $\lambda(x,T)$ in BaFe$_2$(As$_{1-x}$P$_x$)$_2$. As a proof-of-principle, we made an attempt to fit the available experimental data using our microscopic model. The results of this analysis are shown in Fig. \ref{Fig1}. Finally, we also discuss the anomalies associated with the magnetic QCP in the low-temperature dependence of the heat capacity.   

\section{Mean field analysis}

\subsection{London penetration depth}

We start by discussing low-temperature behavior of the magnetic penetration depth in a multiband two-dimensional superconductor with incidental nodes of the gap structure in the clean limit. This particular analysis can be carried out rather generally without an appeal to a particular microscopic model. The total penetration depth $\lambda$ is given by a sum of contributions from each individual bands labeled by index $i$, namely  $\lambda^{-2}=\sum_i\lambda^{-2}_{i}$. Each individual term in the sum is given by \cite{Kopnin}
\begin{equation}
\lambda^{-2}_{i}=\lambda^{-2}_0T\sum_{\omega_n>0}\int\limits^{2\pi}_{0}d\phi\frac{\Delta^2_i(\phi)}{(\omega^2_n+\Delta^2_i(\phi))^{3/2}}
\end{equation}
where $\Delta_i(\phi)$ is the energy gap function and summation goes over the fermionic Matsubara frequencies $\omega_n=\pi T(2n+1)$ with $n\in\mathbb{Z}$. The normalization factor, $\lambda^{-2}_0=4\pi e^2n_i/m_ic^2$, is an effective zero-temperature penetration depth for a given band with an effective mass $m_i$ and carrier concentration $n_i$. To extract the low-$T$ asymptote of $\lambda_i(T)$ and to single out the contribution of gap nodes we convert the sum into an integral. It can be shown that the following result holds   
 \begin{equation}
\sum_{\omega_n}\frac{T}{(\omega^2_n+\Delta_i^2)^{3/2}}=-\frac{2}{\pi}\int\limits^{\infty}_{|\Delta_i|}\frac{d\omega}{\sqrt{\omega_-}}\frac{d}{d\omega}\left[\frac{\tanh\frac{\omega}{2T}}{\omega_+\sqrt{\omega_+}}\right]
\end{equation}
where $\omega_\pm=\omega\pm|\Delta_i|$. Introducing the dimensionless variable $x=\omega/\Delta_i$ one then finds 
\begin{equation}
\frac{\lambda^{-2}_i(T)-\lambda^{-2}_i(0)}{\lambda^{-2}_0}=-\frac{2}{\pi}\int\limits^{2\pi}_{0}\!\!d\phi\!\!\int\limits^{\infty}_{1}\!\!\frac{dx}{\sqrt{x-1}}
\frac{d}{dx}\left[\frac{\tanh\left(\frac{x\Delta_i(\phi)}{2T}\right)-1}{(x+1)^{3/2}}\right]. 
\end{equation}
Let us now assume that gap $\Delta_i(\phi)$ has $N$ isolated simple nodes at some arbitrary locations on the Fermi surface $\phi_k$, $k=1,\ldots,N$. Provided that nodes are not too close to each other we can expand the gap near each such node $\Delta_i(\phi)\approx\Delta\cdot(\phi-\phi_k)\beta_k$, where $\beta_k$ are some numerical factors.  Then with the exponential accuracy in temperature, $T\ll\Delta$, we can approximate the angular integral as follows: $\int^{2\pi}_{0}d\phi\left[\tanh\left(\frac{x\Delta_i(\phi)}{2T}\right)-1\right]\approx 2\int^{\infty}_{0}d\phi\left[\tanh\left(\frac{x\Delta\beta_k\phi}{2T}\right)-1\right]=-\frac{4T\ln 2}{\Delta_0\beta_kx}$. When performing the remaining $x$ integration we get a numerical factor of $\int^{\infty}_{1}\frac{dx}{\sqrt{x-1}}\frac{d}{dx}\left[\frac{1}{x(x+1)^{3/2}}\right]=-\pi/4$, and as a result find for the relative change 
\begin{equation}
\frac{\delta\lambda(T)}{\lambda_0}=\frac{T}{\Delta}\sum_k\frac{\ln2}{\beta_k}
\end{equation}
It can be readily verified that taking $\Delta_i(\phi)=\Delta\cos(2\phi)$ this analysis reproduces classical results for the penetration depth in the $d$-wave case \cite{HirschfeldGoldenfeld}. For the case of iron-pnictides with sign-changing $s^{\pm}$ pairing scenario one finds fully gapped hole pockets with the gap $\Delta_h$, and angular dependence of the gaps along electron Fermi surfaces $\Delta_e(\phi)=\Delta(1-\alpha\cos2\phi)$ \cite{Mazin}.  For $\alpha>1$ there are four incidental nodes counted from $\phi_0=\frac{1}{2}\arccos(1/\alpha)$. This gives correspondingly 
\begin{equation}\label{T-linear-MF}
\frac{\delta\lambda}{\lambda_0}=s\frac{T}{\Delta},\quad s=\frac{2\ln2}{\sqrt{\alpha^2-1}}, 
\end{equation}
which is supposed to hold provided that $\alpha$ is not too close to 1. From the random-phase-approximation based studies of the five-orbital Hubbard model \cite{Graser,Maier} and renormalization-group  analysis of itinerant models \cite{CVV} we can infer $\alpha\approx2-2.5$. The slope $s$ of the relative change in the penetration depth in Eq. \eqref{T-linear-MF} gives us a reference point of the mean-field analysis.  

\subsection{Heat capacity}

We also very briefly discuss the effect of incidental nodes on the low-temperature dependence of the specific heat $C(T)$. The generic thermodynamic formula reads 
\begin{equation}
C(T)=\nu_F\int\limits^{+\infty}_{-\infty}d\xi\int\limits^{2\pi}_{0}\frac{d\phi}{2\pi}E_\phi\frac{\partial }{\partial T}\left[\frac{1}{e^{E_\phi/T}+1}\right]
\end{equation}
where as usual the full integration over the 2D momentum was replaced by the $\xi$-integral over the electronic states close to the Fermi surface with the density of states $\nu_F$ and $E_\phi=\sqrt{\xi^2+\Delta^2(\phi)}$. For a fully gapped Fermi surface one finds exponential suppression of the heat capacity, $\propto e^{-\Delta/T}$, in complete analogy with the classical example of $s$-wave SCs. For a nodal case, with the gap function $\Delta_e(\phi)$, we expand quasiparticle spectrum near the nodal points, $E_\phi\approx\sqrt{\xi^2+(\beta_k\Delta\delta\phi)^2}$, and introduce two dimensionless integration variables $x=\xi/T$ and $y=\beta_k\Delta\delta\phi/T$, where integration over $y$ can be expanded to infinity with the exponential accuracy in $T$. Finally passing to polar coordinates with $r=\sqrt{x^2+y^2}$ one finds for $T\ll\Delta(T)$
\begin{equation}\label{C-MF}
C(T)=\frac{9\zeta(3)}{\sqrt{\alpha^2-1}}\frac{\nu_FT^2}{\Delta},    
\end{equation}
where the numerical factor came from the integral over the radial variable  $\int^{\infty}_{0}r^3dr/\cosh^2(r/2)=18\zeta(3)$. We note that in the three-dimensional case, the same calculation gives $T^3$ due to the phase space integral over the solid angle and recall that for the gap structure with the line nodes it remains that $C\propto T^2$.  

\section{Quantum fluctuation corrections}

\subsection{Band model and SDW propagator} 

Motivated by experimental results \cite{Hashimoto1,Hashimoto2,Auslaender,Joshi}, we proceed with the analysis of quantum fluctuation effects on the London penetration depth. This problem has recently attracted a considerable interest \cite{Levchenko,Chowdhury1,Nomoto,Chowdhury2,Dzero,Huang,Khodas,MDAL}. Our key result concerns the progressive renormalization of the slope in Eq. \eqref{T-linear-MF} of the linear-$T$ dependence of the penetration depth near hidden QCP. 
This result is a direct consequence of the analytical form of the soft-mode propagator inside a superconducting state. Thus the main conclusion is independent of particular microscopic details. However, for concreteness, it is instructive to derive such a propagator from a well-defined microscopic model.

For this purpose, we adopt a relevant three-band model developed in the context of multiband superconductivity in iron-pnictides. Following the earlier works \cite{FS,VVC}, we consider the Hamiltonian that includes the free-fermion part describing  two elliptical electron-like Fermi surfaces and one circular hole-like Fermi surface. The pair-fermion interactions include superconducting and magnetic channels. We assume that the most relevant interaction is between hole and electron pockets, separated by $\bm{Q}=(\pi,\pi)$ in the folded Brillouin zone, and that the gap has $s^\pm$ symmetry, namely it changes sign between electron and hole bands. In the mean-field theory, the phase diagram of this model has been carefully studied. Specifically, with the change in the ellipticity and size of the Fermi surfaces the ground state of the model described above corresponds to the one in which the SC order emerges gradually, and its appearance does not destroy SDW order. In fact, there exists a fairly broad parameter range where SDW and SC orders coexist. Furthermore, the region of coexistence is separated from the pure SC state by a line of the second order phase transition that terminates at $T=0$ magnetic QCP.  

We focus on the low-temperature region of the phase diagram where the system has a long-range SC order and is about to develop magnetic SDW order. In order to capture the main properties of the interplay between SDW and SC, it is convenient to further simplify the model and focus on the interaction between the hole pocket and only one of the two electron pockets. Under these assumptions, the spin-fluctuation propagator is given by
\begin{equation}\label{L}
L(Q,\Omega_m)=\left(g^{-1}_{\text{sdw}}+\Pi(Q,\Omega_m)\right)^{-1}
\end{equation}
where $g_{\text{sdw}}$ is the bare interaction coupling constant in the magnetic channel and the polarization operator is given by 
\begin{equation}\label{Pi}
\Pi=\nu_FT\sum_{\omega_n}\int\limits^{\infty}_{-\infty} d\xi\left\langle\frac{(i\omega_+-\xi_+)(i\omega_-+\xi_-)-\Delta_e\Delta_h}
{(\xi^2_++\omega^2_++\Delta^2_h)(\xi^2_-+\omega^2_-+\Delta^2_e)}\right\rangle_\phi,
\end{equation}
which captures both normal ($GG$) and anomalous ($FF$) contributions. Here $\omega_\pm=\omega_n\pm\Omega_m/2$, $\Omega_m=2\pi mT$, $\xi_\pm=\xi\pm(\varepsilon_\phi+\varepsilon_Q)$, and angular averaging $\langle\ldots\rangle=\int^{2\pi}_{0}\frac{d\phi}{2\pi}(\ldots)$. The band parameter $\varepsilon_\phi=\varepsilon_0+\varepsilon_2\cos(2\phi)$ describes changes in the Fermi surfaces radii and overall shape (ellipticity) induced by doping the system. The second band term, $\varepsilon_Q=(v_FQ/2)\cos(\phi-\psi)$, describes the relative shift in the centers of Fermi surfaces, where $\phi$ and $\psi$ are the directions of $\bm{k_F}$ and $\bm{Q}$. The magnetic SDW quantum critical point is determined in terms of doping parameters $\varepsilon_0$ and $\varepsilon_2$ from the condition that the propagator in Eq. \eqref{L} has a pole at $Q,\Omega_m\to0$, while the region of SDW-SC coexistence persists for $0.8\lesssim\varepsilon_2/\varepsilon_0\lesssim4.7$ \cite{VVC}. 
 
At this point we introduce dimensionless parameter $\Gamma=\nu^{-1}_F(g^{-1}_{\text{sdw}}+\Pi(0,0))$ that measures proximity to a quantum critical point in this model . After evaluating the $\xi$ integral in Eq. \eqref{Pi} we arrive at 
\begin{equation}\label{Gamma} 
\Gamma=\ln\frac{T}{T_{s}}-2\pi T\sum_{\omega_n>0}\left\langle \frac{(E_h+E_e)(E_eE_h+D^2)}{E_hE_e\big[(E_h+E_e)^2+4\varepsilon^2_\phi\big]}-\frac{1}{\omega_n}\right\rangle_\phi
\end{equation}
where $E^2_{e,h}=\omega^2_n+\Delta^2_{e,h}$ and $D^2=\omega^2_n+\Delta_e\Delta_h$. In this expression the coupling constant was absorbed into the definition of the transition temperature into the uncontaminated SDW state $T_s=(2e^{\gamma_E}/\pi)\Lambda e^{-1/\nu_Fg_{\text{sdw}}}$ with the respective ultraviolet cutoff $\Lambda$. We extend this analysis for finite excitation energies, $(v_FQ,\Omega_m)\ll\Delta$, and expand the polarization operator to the leading order. This gives us the propagator of the fluctuating SDW order in the form  
\begin{equation}
L(Q,\Omega_m)=\frac{\nu^{-1}_F}{\Gamma+(Q/Q_c)^2+(\Omega_m/\Omega_c)^2}
\end{equation}
The expansion coefficients $Q_c,\Omega_c$ are model-specific and can be expressed in terms of Matsubara sums similar to the one appearing in Eq. \eqref{Gamma}. For instance, for equal gaps without nodes, $|\Delta_{e,h}|=\Delta$, it is possible to obtain closed analytical expressions 
\begin{equation}
\Omega^{-2}_c=\frac{1}{\Delta^2}\langle f_\Omega(\varepsilon_\phi/\Delta)\rangle_\phi, \quad Q^{-2}_c=\frac{v^2_F}{\Delta^2}\langle f_Q(\varepsilon_\phi/\Delta)\rangle_\phi,
\end{equation}
where dimensionless functions in the $T\to0$  limit take the form 
\begin{align}
&f_\Omega(z)=\frac{1}{8}\left[\frac{1}{1+z^2}+\frac{\arccosh\sqrt{1+z^2}}{|z|(1+z^2)^{3/2}}\right],\\
&f_Q(z)=\frac{\cos^2(\phi-\psi)}{8}\left[\frac{2-z^2}{(1+z^2)^2}-\frac{3|z|\arccosh\sqrt{1+z^2}}{(1+z^2)^{5/2}}\right].
\end{align}
In general $\Gamma,Q_c,\Omega_c$ have a complex functional dependence on $\Delta_{e,h}$ and $\varepsilon_{0,2}$. However, in applications to the calculation of thermodynamic properties, it will be sufficient to simply keep them as phenomenological parameters in a given model as their particular form does not influence statements concerning the temperature dependence of the London penetration depth. 

\subsection{Magnetic penetration depth near QCP} 

We turn our attention to the anomalies in the magnetic penetration depth associated with the QCP. 
We use the Kubo formula to express the fluctuation correction to $\lambda=\lambda_0+\delta\lambda_{\text{QCP}}$, through the correction to the
static, long wavelength limit of the current correlation function, $K = K_0 + \delta K_{\text{QCP}}$, 
\begin{equation}\label{dL-dK}
\frac{\delta\lambda_{\text{QCP}}}{\lambda_0}=-\frac{\delta K_{\text{QCP}}}{K_0},\quad K_0=\frac{1}{2}\nu_Fe^2v^2_F.
\end{equation}
We identify that leading contributions to $\delta K_{\text{QCP}}=\delta K_{\text{DOS}}+\delta K_{\text{MT}}$ stem from the effective mass renormalization captured by the self-energy in the density of states type diagram ($\delta K_{\text{DOS}}$) and from the vertex correction due to the quantum interference Maki-Thompson diagram $(\delta K_{\text{MT}})$. The sum of these terms can be conveniently expressed as follows: 
\begin{equation}\label{dK}
\frac{\partial\delta K_{\text{QCP}}}{\partial\Gamma}=\frac{3}{2}e^2v^2_FT\sum_{\Omega_m}\int\frac{d^2Q}{4\pi^2}\frac{\partial L(Q,\Omega_m)}{\partial\Gamma}F_{l}
\end{equation}
where $F_{l}=F_{\text{DOS}}+F_{\text{MT}}$ is the fermionic loop integral comprised of the trace over the product of four Green's functions. When evaluating these diagrams we took advantage of an important simplification suggested by the separation of energy scales.  Indeed, from the structure of the propagator in Eq. \eqref{L} we see that the mass of soft paramagnons is set by the QCP gap $\Delta_{\text{QCP}}\sim\Delta\sqrt{\Gamma}$, since parametrically $\Omega_c\sim \Delta$. This means that for the most relevant low-energy excitations  $(v_FQ,\Omega_m)\sim \Delta_{\text{QCP}}\ll\Delta$ the fermionic Green's functions in the electromagnetic response kernel $\delta K_{\text{QCP}}$ can be taken at zero boson energy and momentum $Q,\Omega_m\to0$. In the end this implies that fermionic and bosonic Matsubara sums and momentum integrals factorize in the general expression for $\delta K_{\text{QCP}}$. This becomes clear while examining Eq. \eqref{dK} where fermion loop contribution $F_l$ appears as an overall multiplicative factor.  Finally, the reason for differentiation of $\delta K_{\text{QCP}}$ over $\Gamma$ is a matter of practical convenience to extract the leading singular behavior by making integration over the boson energies and momenta convergent at the ultraviolet. Evaluating the Matsubara sum over $\Omega_m$ we find 
\begin{equation}\label{dKQCP}
\frac{\partial}{\partial\Gamma}\left(\frac{\delta K_{\text{QCP}}}{K_0}\right)=\int\frac{3\Omega^4_cF_ld^2Q}{16\pi^2\nu^2_FE^3_Q}\left[\coth\left(\frac{E_Q}{2T}\right)+\frac{E_Q/2T}{\sinh^2\left(\frac{E_Q}{2T}\right)}\right ],
\end{equation}  
where $E_Q=\Omega_c\sqrt{\Gamma+(Q/Q_c)^2}$ and in the immediate vicinity of QCP we can relate $\Gamma$ to the linear deviation from the optimal doping $\Gamma=\gamma|x-x_{\textrm{QCP}}|$. Here for simplicity we take $\gamma$ to be an $x$ independent constant. In Eq. \eqref{dKQCP} it is useful to separate the zero-temperature term explicitly, $\delta K_{\text{QCP}}=\delta K_{\text{Q}}+\delta K_{\text{T}}$, so that all thermal effects are captured by the second term. In the $T\to0$ limit only the first term in the square brackets of Eq. \eqref{dKQCP} contributes, so that evaluating the remaining momentum integral we arrive at   
\begin{equation}\label{dKQ}
\frac{\delta K_{\text{Q}}}{K_0}=
\frac{3\Omega_cQ^2_cF_{\text{l}}}{4\pi\nu^2_F}\sqrt{\Gamma}\propto  \sqrt{|x-x_{\text{QCP}}|}.
\end{equation}
The fact that we got square-root dependence of this correction as a function of detuning from the QCP is not accidental. Indeed, based on precise quantum field theory results of critical phenomena \cite{ZZ} it is known that superfluid density should scale as $A_\pm |x-x_{\text{QCP}}|^{3\nu-1}$ as $x-x_{\text{QCP}}\to\pm0$, where $\nu$ is the critical exponent of the correlation length $\xi\propto|x-x_{\text{QCP}}|^{-\nu}$; furthermore, the ratio of prefactors $A_+/A_-$ is universal. As Eq. \eqref{dKQ} is perturbative at the one-loop level, we should simply use the critical exponent of the Gaussian fixed point $\nu=1/2$ in the generic scaling formula, $|x-x_{\text{QCP}}|^{3\nu-1}$, to verify power-law dependence in Eq. \eqref{dKQ}. As a result, the London penetration depth across QCP is expected to display singular behavior with cusp nonanalyticity. 

\begin{figure}[t!]
  \centering
 \includegraphics[width=3.25in]{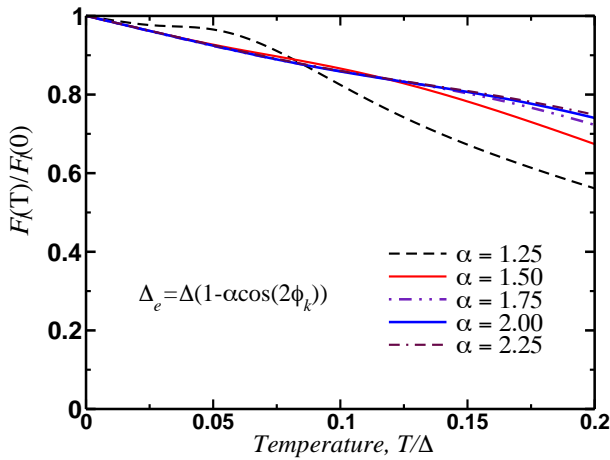}
 \caption{Temperature dependence of the fermionic loop $F_l(T)$ in the electromagnetic response function normalized to its value at $T\to0$ and computed numerically for several different values of the modal parameter $\alpha$ in the gap structure of electronic pockets $\Delta_e(\phi)$. }\label{Fig2}
\end{figure}

We turn our attention to the thermal term now. It is clear from Eq. \eqref{dKQCP} that at lowest temperatures, $T\ll \Delta_{\text{QCP}}$, fluctuation correction $\delta K_{\text{T}}$ is exponentially small, $\propto e^{-\Delta_{\text{QCP}}/T}$, as all degrees of freedom are frozen out. On the other hand, at the intermediate low temperatures, $\Delta_{\text{QCP}}\lesssim T\ll\Delta$, which is still deep SC subgap regime, paramagnon modes already thermally populated 
while quasiparticle states are still gapped out. We have numerically verified that in this limit fermionic loop $F_l(T)$ is very weakly temperature dependent; see Fig. \ref{Fig2} for the illustration.  As a result, provided $T>\Delta_{\text{QCP}}$, we expand over $E_Q/T\ll1$ and integrate over momenta. This gives 
\begin{equation}\label{dKT}
\frac{\delta K_{\text{T}}}{K_0}=-\frac{3TF_lQ^2_c}{4\pi\nu^2_F}\ln\frac{1}{\Gamma}.
\end{equation}
Finally, using expressions for $Q_c$, $F_l=(\nu_F/\Delta^2)\langle f_F(\varepsilon_\phi/\Delta)\rangle_\phi$, and Eq. \eqref{dL-dK} we arrive at the expression for the correction to the penetration depth from quantum critical fluctuations:
\begin{equation}\label{delta-lambda-QCP}
\frac{\delta\lambda_{\text{QCP}}}{\lambda_0}=c_\alpha\frac{T}{E_F}\ln\left(\frac{1}{|x-x_{\textrm{QCP}}|}\right).
\end{equation}
Here we introduced Fermi energy $E_F=\pi\nu_Fv^2_F/4$ and numerical prefactor $c_\alpha=(3/32)(\langle f_F(\varepsilon_\phi/\Delta)\rangle_\phi/\langle f_Q(\varepsilon_\phi/\Delta)\rangle^2_\phi)$. This result implies that the slope of $T$-linear behavior in Eq. \eqref{T-linear-MF} is enhanced near QCP, namely $\delta s_{\text{QCP}}\sim (\Delta/E_F)\ln|x-x_{\textrm{QCP}}|^{-1}$. This trend can be deduced from the data presented in Fig. \ref{Fig1}. 

The two corrections \eqref{dKQ} and \eqref{dKT} have a very different status.
While the zero temperature correction \eqref{dKQ} is peaked around $x=x_{\text{QCP}}$  it does not imply the peak in the penetration depth. 
Instead, the $T$-linear correction \eqref{dKT} implies such a peak \cite{Khodas}.

\subsection{Heat capacity near QCP} 

Motivated by measurements \cite{Carrington} and in a similar fashion to the above analysis, one can account for the effect of fluctuations on the heat capacity \cite{Kuzmanovski,Carvalho}. Indeed, integrating out soft magnetic modes we find a correction to the free energy   
\begin{equation}
\delta F_{\text{QCP}}=\frac{3}{2}T\sum_{\Omega_m}\int\frac{d^2Q}{4\pi^2}\ln\left[L^{-1}(Q,\Omega_m)\right].
\end{equation}
After regularization one finds the corresponding correction to a specific heat, $\delta C_{\text{QCP}}=-T(\partial^2\delta F_{\text{QCP}}/\partial T^2)$, whose leading temperature dependence above the QCP gap, $\Delta_{\text{QCP}}\lesssim T\ll\Delta$, is of the form 
\begin{equation}
\delta C_{\text{QCP}}=\frac{9\zeta(3)}{\pi}\left(\frac{v_FQ_c}{\Omega_c}\right)^2\left(\frac{T}{v_F}\right)^2
\end{equation} 
Similar to the penetration depth, $\delta C_{\text{QCP}}$ is exponentially suppressed $\propto e^{-\Delta_{\text{QCP}}/T}$ at lowest temperatures $T\ll\Delta_{\text{QCP}}$. It is of interest to note that $\propto T^2$ in $\delta C_{\text{QCP}}$ has the same temperature dependence as specific heat in Eq. \eqref{C-MF}. However, in the context of QCP physics $T^2$ is robust and will persist even in the case of fully gapped Fermi surfaces \cite{Khodas}; in other words, nodes are not a prerequisite to have a power law. The relative magnitude of the correction  can be easily estimated $\delta C_{\text{QCP}}/C\sim \Delta/E_F$, since parametrically $v_FQ_c/\Omega_c\sim1$. It should be obviously expected that the same Ginzburg number, $\text{Gi}\sim\Delta/E_F$, controls both perturbative corrections $\delta\lambda_{\text{QCP}}$ and $\delta C_{\text{QCP}}$. We have further estimated higher-loop corrections. For instance, the two-loop contribution to the electromagnetic response kernel at $T=0$ can be deduced to give a correction $\delta K_{\text{Q}}/K_0\propto \text{Gi}^2/\sqrt{x-x_{\text{QCP}}}$. It overcomes the first order term given by Eq. \eqref{dKQ} only inside the Ginzburg region where perturbative analysis breaks down. Thus, with the logarithmic accuracy, the peak height in the penetration depth as estimated from Eq. \eqref{delta-lambda-QCP} is of the order $(\delta\lambda/\lambda_0)_{\text{peak}}\propto (T/E_F)\ln(E_F/\Delta)$.  

\section{Summary and discussion}

In this work we considered the interplay of spin-density wave and superconducting instabilities in multiband metals that lead to the magnetic quantum critical point hidden inside of the superconducting state. 
Magnetic fluctuations are shown to influence thermodynamic properties of the system. While nodal structure of the gap gives power-law temperature dependencies of the magnetic penetration depth and specific heat, fluctuations result in strong renormalizations that are most pronounced in parts of the phase diagram in proximity to the QCP. 

In the context of the penetration depth we expect a peak feature as a function of doping across QCP. Furthermore, the peak is expected to be asymmetric. This can be understood simply by realizing that the spectrum of collective magnetic excitations is different in the ordered and disordered phases. In the paramagnetic phase all three spin polarization directions contribute equally to the thermodynamic properties. In contrast, in the ordered phase, one mode (longitudinal) becomes massive, while two other (transverse) spin fluctuations are turned into massless (Goldstone) modes. Thus a different number of the soft modes contributes to thermodynamic properties. 

Finally, the key observation we draw from the calculation is that having nodal gap structure is not a requirement for a power law in temperature dependence of thermodynamic variables induced by quantum fluctuations. This effect persists even in the scenario of fully gapped Fermi surfaces. For this reason it is also insensitive to the disorder scattering that tends to suppress unconventional symmetry of the superconducting gap structure. 

\section*{Acknowledgments} 

We thank E. Berg, V. S. de Carvalho, D. Chowdhury, A. Chubukov, R. Fernandes, S. Gazit, D. Orgad, R. Prozorov, S. Sachdev, J. Schmalian, and M. Vavilov for numerous useful discussions on the topics related to this study. The experimental data presented in Fig. \ref{Fig1} was made available to us by courtesy of T. Shibauchi and Y. Matsuda per Ref. \cite{Hashimoto1}. 

The work on this project at UW-Madison (A.L.) was financially supported by the U.S. Department of Energy (DOE), Office of Science, Basic Energy Sciences (BES) Program for Materials and Chemistry Research in Quantum Information Science under Award No. DE-SC0020313. M.D. acknowledges financial support from the National Science Foundation Grant No. NSF-DMR-BSF-2002795. M.K. acknowledges financial support from the Israel Science Foundation, Grant No. 2665/20, and in part by the BSF Grant No. 2016317. This work was performed in part at Aspen Center for Physics, which is supported by National Science Foundation Grant No. PHY-1607611.

\end{document}